\documentclass[aps,prl,twocolumn,preprintnumbers,groupedaddress]{revtex4}
\usepackage{amssymb}
\usepackage{epsfig} 
\usepackage{amsmath,color} 
\usepackage{graphicx} 
\usepackage{ulem}

\usepackage{wrapfig}

\newcommand{\beq}{\begin{equation}}
\newcommand{\eeq}{\end{equation}}
\newcommand{\bea}{\begin{eqnarray}}
\newcommand{\eea}{\end{eqnarray}}
\newcommand{\ba}{\begin{array}}
\newcommand{\ea}{\end{array}}
\newcommand{\bi}{\begin{itemize}}
\newcommand{\ei}{\end{itemize}}
\newcommand{\bn}{\begin{enumerate}}
\newcommand{\en}{\end{enumerate}}
\newcommand{\bc}{\begin{center}}
\newcommand{\ec}{\end{center}}
\renewcommand{\l}{\left}
\renewcommand{\r}{\right}

\newcommand{\eq}[1]{Eq.~(\ref{#1})}
\newcommand{\eqs}[2]{Eqs.~(\ref{#1}) and (\ref{#2})}

\newcommand{\GeV}{\mathinner{\mathrm{GeV}}}
\newcommand{\TeV}{\mathinner{\mathrm{TeV}}}

\begin{document}

\preprint{FTUV-15-07-31}
\preprint{IFIC-15-49}

\title{Peccei-Quinn field for inflation, baryogenesis, dark matter, and much more}


\author{Gabriela Barenboim}
\email[]{Gabriela.Barenboim@uv.es}%{Your e-mail address}
\author{Wan-Il Park}
\email[]{Wanil.Park@uv.es}%{Your e-mail address}
%\thanks{}
\affiliation{
Departament de F\'isica Te\`orica and IFIC, Universitat de Val\`encia-CSIC, E-46100, Burjassot, Spain}


\date{\today}

\begin{abstract}
We propose a scenario of brane cosmology in which the Peccei-Quinn field plays the role of the inflaton and solves simultaneously many cosmological and phenomenological issues such as the generation of a heavy Majorana mass for the right-handed neutrinos needed for seesaw mechanism, MSSM $\mu$-parameter, the right amount of baryon number asymmetry and dark matter relic density at the present universe, together with an axion solution to the strong CP problem without the domain wall obstacle. 
Interestingly, the scales of the soft SUSY-breaking mass parameter and that of the breaking of $U(1)_{\rm PQ}$ symmetry are lower bounded at $\mathcal{O}(10) \TeV$ and $\mathcal{O}(10^{11}) \GeV$, respectively.
\end{abstract}

%\keywords{}

\maketitle

%\section{Introduction}
%\subsection{}
%\subsubsection{}

\section{Introduction}

Probably, the most compelling source for the density perturbations in the present universe would be the classicalized quantum fluctuations of the inflaton \cite{Mukhanov:1981xt,Mukhanov:1982nu}.
However, in addressing the spectral features of the density perturbations observed by Planck satellite \cite{Ade:2015xua}, the flatness problem (called $\eta$-problem) of the inflaton potential \cite{Copeland:1994vg,Stewart:1994ts} turns out to be a big obstacle in the attempts of a simple realization of inflation based on an UV theory, e.g., supergravity or string theories with Einstein gravity. 

Actually, the $\eta$-problem can be removed rather easily if the  inflaton path is a simple compactified trajectory in a multi-dimensional field space without resorting to trans-Planckian excursions for the inflaton field \cite{Silverstein:2008sg,McAllister:2008hb,Berg:2009tg,McDonald:2014oza,McDonald:2014nqa,Li:2014vpa,Carone:2014cta,Barenboim:2014vea,Barenboim:2015zka,Barenboim:2015lla} (see also \cite{Kim:2004rp}).
%
%may arise because a single inflaton field is demanded to be responsible for the density perturbations. 
%However, this should not be necessarily the case, since there are many scalar fields in high energy theories and inflation can be just a trajectory in an otherwise multi-dimensional field space.
%For example, recently, there have been interesting ideas on realizing large field inflation in a sub-Planckian field space \cite{Silverstein:2008sg,McAllister:2008hb,Berg:2009tg,McDonald:2014oza,McDonald:2014nqa,Li:2014vpa,Carone:2014cta,Barenboim:2014vea,Barenboim:2015zka,Barenboim:2015lla} (see also \cite{Kim:2004rp}).
%The key of these scenarios is compactifying the trans-Planckian trajectory of the inflaton by winding (or spiraling) it  in a two-dimensional field space.
Notably, the resulting inflation is effectively the same as single-field slow-roll inflation.

Another interesting possibility for tackling the $\eta$-problem is to introduce an extra-dimension \cite{Randall:1999vf}.
%In this case, we are supposed to live on a brane located at a point along the 4-th spacial dimension.
Depending on the relative size of the energy density on the brane as compared with the brane tension, the expansion rate on the brane can be much larger than the one expected in Einstein gravity.
This can resolve the $\eta$-problem, allowing inflation even with a steep potential \cite{Copeland:2000hn}.

In this letter, we show that the  Peccei-Quinn field \cite{Peccei:1977hh} of the DFSZ axion model \cite{Zhitnitsky:1980tq,Dine:1981rt} in the supersymmetric framework is a natural candidate for the inflaton which can also trigger some necessary post-inflation cosmology such as baryogenesis and dark matter,when considering brane cosmology.

\section{Inflation along a flat direction on a brane}
In Randall-Sundrum type II brane world scenario \cite{Randall:1999vf}, after the bulk contribution to the brane is diluted away, Friedman equation on the brane can be written as \cite{Flanagan:1999cu}
\beq
3 H^2 M_{\rm P}^2 = \rho \l( 1 + \frac{\rho}{\Lambda} \r)  
\eeq
where Randall-Sundrum condition for vanishing cosmological constant is assumed, $H$ is the expansion rate on the brane, $M_{\rm P}=2.4 \times 10^{18} \GeV$ is the reduced Planck mass, and $\rho$ is the energy density on the brane other than the brane tension $\Lambda/2$ which is constrained to be larger than about $\l( 2.3 \TeV \r)^4$ from the validity of the $(3+1)$-dimensional Newtonian gravity on scales larger than about $44 \mu{\rm m}$ \cite{Chung:2000rg,Adelberger:2009zz}.
It implies that, if $\tilde{V} \equiv \rho/\Lambda \gg 1$ with $\rho$ dominated by the potential energy $V$ of a scalar field, a long enough inflation is possible even with a steep potential
which would not allow or sustain an inflationary epoch in Einstein gravity.
The slow-roll parameters of such an inflation on the brane are given by \cite{Maartens:1999hf}
\bea
\epsilon &\equiv& - \frac{\dot{H}}{H^2} \approx \epsilon_E \frac{1+ 2 \tilde{V}}{(1+\tilde{V})^2}, 
\\
\eta &\equiv& \frac{V''}{3 H^2} \approx \frac{\eta_E}{1+\tilde{V}}, 
\\
\xi^2 &\equiv& \frac{V' V'''}{\l( 3 H^2 \r)^2} \approx \frac{\xi^2_E}{(1+\tilde{V})^2} 
\eea
where $\epsilon_{\rm E}, \eta_{\rm E}$ and $\xi_{\rm E}^2$ are the conventional slow-roll parameters in Einstein gravity.
The spectral index and its running can be expressed as 
\bea
n_s &=& 1 - 6 \epsilon + 2 \eta, 
\nonumber \\
\frac{d n_s}{d \ln k} &=& 16 \epsilon \eta - 24 \epsilon^2 \frac{1+3\tilde{V}+3\tilde{V}^2}{(1+2 \tilde{V})^2} - 2 \xi^2
\eea
Notice that the spectral index is given by the same expression  as in the case of Einstein gravity.
%The power spectrum is given by
%\beq
%\bar{P}_\mathcal{R} = \l( 1+\tilde{V} \r)^3 P_\mathcal{R} = \frac{\l( 1+\tilde{V} \r)^3}{\epsilon} \frac{V}{24 \pi^2 M_{\rm P}^4}
%\eeq
The power spectrum and  tensor-to-scalar ratio yield \cite{Maartens:1999hf}
\beq
P_\mathcal{R} = \l( 1+\tilde{V} \r)^3 P_{\mathcal{R}, \rm E}, \quad r_T = \frac{16 \epsilon}{1+2\tilde{V}}
\eeq
with $P_{\mathcal{R}, \rm E} = V / (24 \pi^2 M_{\rm P}^4 \epsilon_{\rm E})$, while the number of $e$-foldings is
\beq \label{Ne-th}
N_e = \int H dt = - \frac{1}{M_{\rm P}^2} \int \frac{V}{V'} \l( 1+ \tilde{V} \r) d \phi
\eeq
with $\phi$ being the inflaton field.
For the horizon scale of the present universe $k_H^{-1} \sim 6000 {\rm Mpc}$, the $e$-foldings expanded at the time the Planck pivot scale ($k_* = 0.05 {\rm Mpc}^{-1}$) exits the horizon during inflation becomes \cite{KT}
\bea \label{Ne-obs}
N_{e,*} &\approx& \frac{1}{3} \ln \l[ \frac{\sqrt{6}}{16 \pi} \mathbb{S}_0 \l( \frac{k_0}{k_*} \frac{V_*}{V_e} \frac{V_e^{\frac{1}{4}}}{M_{\rm P}} \r)^3 \l( \frac{T_{\rm d}}{\Lambda^{\frac{1}{4}}} \r) \tilde{V}_e^{\frac{5}{4}} \r]
\nonumber \\
&\simeq& 39.4 - \ln \l[ \frac{V_e}{V_*} \l( \frac{10^{8} \GeV}{V_e^{\frac{1}{4}}} \r) \r] - \frac{1}{3} \ln \l[ \frac{\Lambda^{\frac{1}{4}}}{T_{\rm d}} \r]
\nonumber \\
&& \phantom{39.4} + \frac{5}{12} \ln \tilde{V}_e
\eea
where $\mathbb{S}_0$ is the total entropy of the present observable universe, the subscript `$_e$' in the right-hand side denotes a value at the end of inflation, and $T_{\rm d}$ is the decay temperature of the inflaton. 
Note that the last term in the second line of \eq{Ne-obs} is the new contribution coming from the enhancement of the expansion rate.
Also, in order not to over-produce KK-gravitions, $T_{\rm d} \lesssim (\Lambda/2)^{1/4}$ is demanded \cite{Allahverdi:2001eq}. 
%\beq
%\bar{N}_{e,*} 
%= 56.3 + \frac{1}{2} \ln \l( 1 + \tilde{V} \r) - \ln \frac{10^{16} \GeV}{V_I^{1/4}} - \frac{1}{12} \ln \frac{V_I}{\rho_{\rm R}}
%\eeq
%where the second term of the right-hand side is the new contribution coming from the enhancement of the expansion rate, $V_I$ is the energy density of inflation, and $\rho_{\rm R}$ is the radiation density at reheating. 

Generically, thanks to the non-renormalization theorems, the potential $V$ of a supersymmetric flat direction ($\Phi = \phi e^{i \theta}/\sqrt{2}$) is dominated by a mass term until the field gets close to the true vacuum or before it is lifted up by higher order terms.
For a symmetry-breaking flat direction, the potential can be written as
\beq
V = V_0 - \frac{1}{2} m^2 \phi^2 + \dots
\eeq
where $V_0$ is set for vanishing cosmological constant at the true vacuum, $m^2(>0)$ is a mass-square parameter, and `$\dots$' represents higher order terms suppressed by a very large mass scale, e.g., Planck scale.
Roughly, $V_0 \sim m^2 \phi_0^2$ with $\phi_0$ being the vacuum expectation value (VEV).
Then, for $\phi \ll \phi_0$ one finds 
\beq
\frac{M_{\rm P}^2 V''}{V} \sim - \l( \frac{M_{\rm P}}{\phi_0} \r)^2, \ \l| \frac{M_{\rm P} V'}{V} \r|^2 \sim \l( \frac{\phi}{\phi_0} \r)^2 |\eta|
\eeq
%and
%\beq
%\epsilon \simeq \frac{1}{2} \l( \frac{\phi}{M_{\rm P}} \r)^2 \eta^2 \sim \frac{1}{2} \l( \frac{\phi}{\phi_0} \r)^2 |\eta|
%\eeq
%Hence, for $\tilde{V} \gg 1$, 
%\beq
%\bar{\epsilon} \simeq \frac{2 \epsilon}{\tilde{V}} \sim \l( \frac{\phi}{\phi_0} \r)^2 |\bar{\eta}|
%\eeq
%and for $\phi_* \ll \phi_0$ which will be justified shortly a slow-roll inflation on a brane results in
%Hence, for $V/\Lambda \equiv \tilde{V} \gg 1$ and $\phi_* \ll \phi_0$ which will be justified shortly, one finds $\bar{\epsilon} \ll \bar{\eta}$ and a slow-roll inflation on a brane results in
%Hence, for $V/\Lambda \equiv \tilde{V} \gg 1$ and $\phi_* \ll \phi_0$ which will be justified shortly, the slow-roll parameters of brane inflation are such that $\epsilon \ll \eta$ resulting in
Hence, if $\phi \ll \phi_0$ for a relevant cosmological scale during inflation, the slow-roll parameters of brane inflation are such that $\epsilon \ll \eta$ resulting in
\beq
n_s-1 \approx 2 \eta
\eeq
Combining with the number of $e$-foldings in \eq{Ne-th}, we find 
\beq \label{inf-cond}
N_{e,*} \l( n_s - 1 \r) \sim 2 \ln \l( \frac{\phi_*}{\phi_e} \r)
\eeq
%where the subscript `$_*$' and $\phi_e$ denoting the horizon-exit of a cosmological scale and the value of $\phi$ at the end of inflation.
\eq{inf-cond} implies 
\beq
\phi_* \sim \phi_e e^{N_{e,*} \l( n_s - 1 \r)/2} \sim 0.37 \phi_e = \mathcal{O}(0.1) \phi_0
\eeq
where for a numerical estimation we used $N_{e,*}=50$ and $\eta \approx - 1.7 \times 10^{-2}$ \cite{Ade:2015xua} leading to
\beq \label{tilde-V}
\tilde{V} \sim 60 \l( M_{\rm P}/\phi_0 \r)^2
\eeq

The power spectrum can be expressed as 
\beq
P_\mathcal{R} \sim \frac{1}{12 \pi^2 |\eta|^3} \l( \frac{m}{\phi_*} \r)^2
\eeq
and $P_\mathcal{R} = 2.142 \times 10^{-9}$ which with $\eta = - 1.7 \times 10^{-2}$ \cite{Ade:2015xua} leads to
\beq \label{m-over-phi0}
m/\phi_0 \sim 10^{-7}
\eeq
Also, from \eqs{tilde-V}{m-over-phi0}, the scale of brane tension is found to be
\beq
\l( \Lambda/2 \r)^{1/4} \sim 0.6 \l( \frac{m}{10^5 \GeV} \r)^{1/2} m
\eeq
Since $\Lambda$ is constrained as $(\Lambda/2)^{1/4} \gtrsim 2.3 \TeV$, we find 
\beq
\mathcal{O}(10) \TeV \lesssim m \lesssim m_{\rm s} 
\eeq
with $m_{\rm s}$ being the scale of the soft SUSY-breaking mass parameter, and 
\beq
\phi_0 \gtrsim \mathcal{O}(10^{11}) \GeV
\eeq
It is very interesting to notice that, the Peccei-Quinn field is a natural and compelling candidate of $\phi$.

\section{The model}
Motivated by the observation in the previous section, we consider the Peccei-Quinn field as the inflaton.
For a specific realization,  we consider a model defined by the following superpotential where gauge group and family indices are omitted:
\bea \label{W-full}
W 
&=& Y_u Q H_u \bar{u} + Y_d Q H_d \bar{d} + Y_e L H_d \bar{e}
\nonumber \\
&& + \frac{\lambda_N}{2} X N^2 + \lambda_\nu L H_u N
\nonumber \\
&& + \frac{\lambda_\mu X^2 H_uH_d}{M_{\rm P}} + \frac{\lambda_{XY} X^3 Y}{M_{\rm P}}
\eea
here $U(1)$ Peccei-Quinn symmetry is assumed with charges assigned as $q_{\rm PQ}(X,Y) = (1,-3)$.
The first line of \eq{W-full} is the supersymmetric realization of the standard model (SM) Yukawa couplings.
The second line triggers the seesaw mechanism when $X$ develops a non-zero vacuum expectation value (VEV).
That is, right-handed neutrinos (RHNs) get masses given by
\beq
m_N = \lambda_N \langle X \rangle
\eeq
where the family index was suppressed.
As $m_N \gg H$, RHNs are integrated out, replacing the second line of \eq{W-full} by  
\beq \label{W-seesaw}
- \frac{1}{2} \frac{\lambda_\nu^2 \l( LH_u \r)^2}{\lambda_N \langle X \rangle} 
\eeq 
which generates  Majora masses of active neutrinos at low energy given by
\beq
m_\nu = \frac{\lambda_\nu^2 v_u^2}{\lambda_N X_0}
\eeq
with $v_u$ being the VEV of $H_u$.
The third line of \eq{W-full} is nothing but a supersymmetric realization of the DFSZ axion model \cite{Kim:1983dt}. %(see for example Ref.~\cite{Chun:2000jx} for some details.). 
The first term of the third line reproduces the $\mu$-term of the minimal supersymmetric standard model (MSSM) as $X$ develops a non-zero VEV, and the second term  stabilizes $X$ and $Y$.

For a simple illustration of inflation and post-inflation cosmology in the model of \eq{W-full}, we set $X = Y = \Phi$ and consider a superpotential,
\beq \label{W-PQ}
W = \frac{\lambda}{4} \frac{\Phi^4}{M_{\rm P}}
\eeq
where $\lambda=\mathcal{O}(0.1-1)$ is a dimensionless constant.
Then, including the soft SUSY-breaking terms with a negative mass-square parameter assumed \footnote{The negativity of the  mass-square parameter can be obtained, for example, by a radiative running or specific choice of the mechanism of SUSY-breaking and its mediation.}, the scalar potential of $\Phi$ ($\Phi^*$) becomes 
\beq \label{V-PQ}
V = V_0 - m^2 |\Phi|^2 - \l( \frac{A \lambda}{4} \frac{\Phi^4}{M_{\rm P}} + {\rm c.c.} \r) + \l| \frac{\lambda \Phi^3}{M_{\rm P}} \r|^2
\eeq
where $V_0$ is again set to get a vanishing cosmological constant at true vacuum, and $m^2(>0)$ and $A$ are soft SUSY-breaking parameters.
The vacuum expectation value of $\Phi$ is found to be 
\beq
\Phi_0 = \l( \frac{A M_{\rm P}}{6 \lambda} \r)^{1/2} \l[ 1 + \sqrt{1 + \frac{12 m^2}{A^2}} \r]^{1/2}
\eeq
and
\beq
V_0 = \frac{2}{3} m^2 |\Phi_0|^2 \l[ 1 + \frac{1}{24} \frac{A^2}{m^2} \l( 1 + \sqrt{1 + \frac{12 m^2}{A^2}} \r) \r] 
\eeq
The masses of radial and angular modes at true vacuum are 
\bea \label{r-msq}
m_{\rm PQ, r}^2 &=& 2 m^2
\\ \label{a-msq}
m_{\rm PQ, a}^2 &=& \frac{2}{3} A^2 \l[ 1 + \sqrt{1+ \frac{12 m^2}{A^2}} \r]
\eea

\section{Cosmology}
The model of \eq{W-full} can accommodate inflation in a brane world scenario, and trigger the realization of a successful baryogenesis and supply dark matter production as follows.

\subsection{Inflation}
Although inflation in the model of \eq{W-full} takes place along $X$, all its main features can be captured from the simplified example of \eq{W-PQ}.
From our numerical analysis, we found that, as an example, the following set of parameters gives a perfect fit to the Planck data:
\beq
m \simeq 1.4 \times 10^5 \GeV, \ \lambda \simeq 0.61, \ 2 \tilde{V} = 10^{15.64}
\eeq
which results in
\beq
m_{\rm PQ, r} = 1.96 \times 10^5 \GeV, \ \phi_0 = 9.15 \times 10^{11} \GeV
\eeq
and $\l( \Lambda/2 \r)^{\frac{1}{4}} \simeq 51 \TeV$.
One can also lower $\phi_0$ while keeping $m/\phi_0$ nearly invariant for a correct value of $P_\mathcal{R}$.
The decay rate of the flaton is assumed to be \cite{Kim:2008yu}
\beq \label{Gamma-d}
\Gamma_\phi = \frac{1}{4 \pi} \l( 1 - \frac{B^2}{m_A^2} \r)^2 \l( \frac{\mu}{m_{\rm PQ, r}} \r)^4 \frac{m_{\rm PQ, r}^3}{\phi_0^2}
\eeq
where $B$ is the soft SUSY-breaking parameter of the Higgs bilinear term in MSSM and $m_A$ is the mass of CP-odd Higgs.
For $m_A = 1.03 B$ and $\mu = m$, one finds $T_{\rm d} \simeq 550 \GeV$ leading to $N_{e,*} \simeq 55$.
Inflation ends at $\phi_e \simeq 0.8 \phi_0$, and inflationary observables turn out to be
\beq
10^9 P_\mathcal{R} \simeq 2.136, \ n_s \simeq 0.962, \ \frac{d n_s}{d \ln k} = -6.1 \times 10^{-4}
\eeq
The tensor-to-scalar ratio is miserably small, undistinguishable zero from an experimental point of view.

Even if $\Phi$ evolves in a 2-dimensional field space, for $\phi_* \lesssim \phi \lesssim \phi_e$, the angular direction of $\Phi$ is much flatter than radial direction.
Hence, unless the trajectory of the inflaton is around the ridge of an angular potential (e.g., the potential involving $A$-parameter in \eq{V-PQ}), the non-Gaussianities are expected to be $f_{\rm NL} \ll 1$, since perturbations are not enough to produce significant changes on the inflaton trajectories which nicely converge to the true vacuum.
This is expected to be true as well for the real model of \eq{W-full} where $X$ plays the role of the slow-roll inflaton.
Note that $Y$ develops its VEV via the $A$-term of the second term in \eq{W-full}.
So, $Y$ is expected to follow its time-dependent vacuum position during the whole period of inflation, and therefore it would not affect the adiabatic density perturbations caused mainly by the perturbations of $X$.

A remark on the initial condition of the inflation along PQ-field is in order.
If the PQ-field were the only light field, we can simply assume a radiation dominant universe before the onset of inflation.
However, there are at least  two possible candidates of light fields too: GUT Higgs and Planckian moduli.
For $\Lambda^{1/4} = \mathcal{O}(10^{3-5}) \GeV$ which is the case for the parameter set we have chosen for illustration purposes, they can also lead inflation while they roll out/in from/to the origin.
As a simple possibility to realize PQ-inflation as the last inflation that may be necessary to have a proper post-inflation cosmology (e.g., baryogenesis without unwanted relic problem), we may assume that the mass scales of those field are slightly higher than that of the PQ-field.
In this case, the expansion rate at the end of the pre-existing inflation ($H_{\rm pre}$) could be a little bit larger than the one from PQ-inflation.
The PQ-field would barely move from the origin once it is there.
The potential danger of being drifted away by quantum fluctuations can be removed if PQ-field couples to a light degree of freedom around the origin.
Specifically, in the presence of a coupling $W \supset \lambda_N \Phi N^2$, the effective mass-square of the PQ-field around the origin is 
\beq
- m^2 + \lambda_N^2 \l( \frac{H_{\rm pre}}{2 \pi} \r)^2
\eeq
since RHNs would have fluctuations of order $H_{\rm pre}/(2 \pi)$.
Hence, if 
\beq
H_{\rm pre} >2 \pi m / \lambda_N
\eeq
the PQ-field can be safely held at the origin.
In other words, the mass scale of GUT Higgs or Planckian moduli is required to be larger than that of PQ-field by a factor about $2 \pi/\lambda_N$.
If GUT Higgs causes pre-inflation, the reheating temperature is likely to be low, since the interaction to light fields will be suppressed by GUT scale.
In this case, the PQ-inflaton may start rolling out soon after the pre-inflation if $H_{\rm pre} \sim 2 \pi m / \lambda_N$. 
On the other hand, Planckian moduli may end up at a symmetry-enhanced point leading to a rapid reheating and resulting in a rather high reheating temperature (e.g., $\sim \l(6 H_{\rm pre}^2 M_{\rm P}^2 \Lambda \r)^{1/8}$).
In this case, the PQ-inflaton could be held at the origin until the temperature of thermal bath becomes comparable to $m$ (a phase of thermal inflation \cite{Lyth:1995hj,Lyth:1995ka}).

Our choice of parameters results in $H \sim \mathcal{O}(10) \l( \Lambda/2 \r)^{1/4}$ during inflation.
In this case, the higher order kinetic terms of the inflaton, caused by brane fluctuations, are not suppressed enough, making the quadratic kinetic term a rather poor approximation \footnote{We thank anonymous referee for pointing this issue out.}.
It is non-trivial to properly handle this issue given the tachyonic nature of our inflaton, and out of the scope of this paper.
However, at least, once the brane tension is given, enforcing the validity of considering only the quadratic kinetic term of the inflaton as we did in this work may impose a constraint on the energy density of the inflation $V_0$ such that Hubble fluctuation is at most of the same order as the brane tension.
As a simple solution to this issue in our scenario, one can increase the size of $A$-parameter relative to the mass parameter $m$ by a factor $\sim \mathcal{O}(10)$ in \eq{V-PQ}.
In this case, the contribution of the $A$-term to the potential for the cosmological scales of interest becomes sizable, requiring a smaller $\phi_*$ in order to obtain the right amount of $e$-foldings.
This means that $\epsilon_*$ becomes smaller.
As a result, for a given brane tension one can lower the expansion rate during inflation by taking a slightly smaller mass scale for both parameters, $m$ and $A$ in \eq{V-PQ}, while keeping inflationary observables nearly unchanged.
Baryogenesis and the production of dark matter discussed in the next subsections are barely affected although there will be minor changes in the working parameter space.  
So, we do not pursue the details of this case here.

\subsection{Baryogenesis}
The dynamical generation of $\mu$ can sustain a late time Affleck-Dine leptogenesis, as discussed in Refs.~\cite{Jeong:2004hy,Kim:2008yu,Park:2010qd}, if 
\beq \label{mLHu}
m_{LH_u}^2 \equiv \l( m_L^2 + m_{H_u}^2 \r)/2 < 0, \ 2 m_{LH_u}^2 + |\mu|^2 > 0
\eeq
at scales below an intermediate scale, where $m_L^2$ and $m_{H_u}^2$ are respectively the soft SUSY-breaking mass-square parameters of $L$ and $H_u$ fields.
In addition to \eq{mLHu}, we also assume
\beq
|m_{LH_u}^2| \gtrsim m^2
\eeq
with $-m^2$ being understood as the soft SUSY-breaking mass-square of $X$ so that the PQ-field rolls away from the origin following the $LH_u$-flat direction.
Note that, in the model of \eq{W-full} when $X$ is held around the origin, the $LH_u$ direction is stabilized by a quartic term.
However, as $X$ rolls out and becomes large, the renormalizable terms in the second line of \eq{W-full} are replaced by the one of \eq{W-seesaw}.
The inverse of the time scale of the evolution of $\langle \ell \rangle$, the time-dependent VEV of the $LH_u$ flat-direction, is 
\beq
\dot{\langle \ell \rangle}/\langle \ell \rangle = \dot{X}/X = - \eta_X H
\eeq
where $\eta_X$ is the slow-roll parameter of the inflaton ($X$).
It is much smaller than the mass scale of $LH_u$ around the potential minimum.
Hence, the $LH_u$ flat-direction would trace its potential minimum as $X$ evolves.   

%One can also consider the case where a seesaw operator exists from the beginning.
%In this case, the $LH_u$ flat-direction and the PQ-field roll out nearly simultaneously, and the solution of the equation of motion can be approximated as
%\beq
%\ell = \ell_i e^{\eta_\ell \bar{H}(t-t_{\rm c})}, \ X = X_i e^{\eta_X \bar{H}(t-t_{\rm c})}
%\eeq
%where $\eta_\ell \equiv m_{LH_u}^2/(3 H^2)$.
%Hence, if we set for simplicity $\eta_\ell = \eta_X$, we find
%\beq
%\Delta N_e \equiv \bar{H}(t_{X,0} - t_{\ell,0}) =\frac{1}{\eta_X} \ln \frac{X_0}{\ell_0}
%\eeq
%In order for our baryogenesis mechanism to work, we need $X_0/\ell_0 \gtrsim \mathcal{O}(10)$.
%Hence, $\Delta N_e$ is much larger than the one associated with the present horizon scale.
%This means that typically, the $LH_u$ flat direction would settle down to its vacuum well before the pivot scale of Plank data exits the horizon during inflation.

The lepton number asymmetry at the onset of the angular motion of the AD field is estimated to be 
\beq \label{nL}
n_L \sim \frac{A_\nu}{m_{\rm PQ,r}} \frac{\lambda_\nu^2 \ell_e^4}{\lambda_N X_0} \sin (4 \theta_\ell + \Delta \theta)
\eeq
with $m_{\rm PQ,r}$ being the mass of the radial mode of $X$ around the true vauum, and $\ell_e$ is the VEV of $LH_u$ flat direction at the end of inflation.
Because the oscillation of $X$  causes time-dependent variations of the mass and of $\ell_e$ of $LH_u$, the initial lepton number asymmetry is suppressed by a factor which we denote as $\delta_{\rm s}$.
%
%$\delta_{\rm s}$ is the suppression factor of the generated lepton number asymmetry relative to the initial number density of AD field.
%
Passing through the electroweak symmetry breaking, the asymmetry in the lepton number is converted to the one in baryon number via anomalous electroweak processes.
Then, taking $n_B \sim n_L/3$, the late time baryon number asymmetry after the inflaton decay can be expressed as
\bea \label{nB-over-s}
\frac{n_B}{s} &\sim& \delta_{\rm s} \frac{T_{\rm d}}{4 m_{\rm PQ,r}} \l( \frac{A_\nu}{m_{\rm PQ,r}} \r)^2 \l( \frac{\ell_0}{X_{\rm osc}} \r)^2 
\nonumber \\
&& \times f(A_\nu) \l( \frac{\ell_e}{\ell_0} \r)^4 \sin (4 \theta_\ell + \Delta \theta)
\eea
where we use
\beq
\ell_0^2 = \frac{A_\nu M_\nu}{6 \lambda_{\bar{\nu}}} \l[ 1 + \sqrt{1+\frac{12 |m_{LH_u}^2|}{A_\nu^2}} \r]
\eeq 
where $M_\nu \equiv \lambda_N X_0$ and $\lambda_{\bar{\nu}} \equiv \lambda_\nu^2$, $X_{\rm osc} = \alpha X_0$ is the initial oscillation amplitude of $X$ at the end of inflation with $\alpha = \mathcal{O}(0.1)$, $f(A_\nu) \equiv \lambda_{\bar{\nu}} \ell_0^2/ \l( A_\nu M_\nu \r)$, and $\Delta \theta$ is the relative CP-violating phase between the two lepton-number violating terms \cite{Park:2010qd} and assumed to be of order one.
The precise estimation of $\delta_{\rm s}$ requires a heavy numerical simulation which is out of the scope of this letter.
However, from the simulation results of Ref.~\cite{Kim:2008yu}, we expect $\delta_{\rm s} = \mathcal{O}(10^{-3}-10^{-2})$, taking into account the larger expansion rate at the time of AD leptogenesis and the weaker preheating due to the smallness of the $\mu$-term-induced mass variation of the $LH_u$ direction.
For the parameter set used in the previous section with $A_\nu \sim \sqrt{|m_{LH_u}^2|} \sim m$, we obtain $\ell_0 \sim 10^{9-10} \GeV$ and the factors appearing in the second line of \eq{nB-over-s} are $\mathcal{O}(0.1)$ (once taken together).
Hence, %it would be possible to obtain the right amount of baryon number asymmetry provided 
the right amount of baryon number asymmetry would always be obtained if 
\beq
T_{\rm d} = \mathcal{O}(10^{-3}-10^{-2}) m
\eeq 
Notice that, unless $\delta_{\rm s}$ is too small, the right amount of baryon number asymmetry can be obtained by adjusting (mostly lowering down) $T_{\rm d}$ (or $B/m_A$ - see \eq{Gamma-d}). 
%Compared to the case discussed in \cite{}, in our scenario the expansion rate at the time of AD leptogenesis is comparable to the expansion rate which rapidly decreases though relative to the case of GR.
%This helps preserving the initially generated lepton number asymmetry.
%However, since the $\mu$-term-induced mass variation of $LH_u$ direction is rather small, compared to the case of Ref.~\cite{}.
%This may suppress the initial lepton number asymmetry since the effect of preheating would be rather weak.
%Hence, we expect the eventual conserved lepton number asymmetry after complicated dynamics would be little bit larger than or more-or-less same as the one in Ref.~\cite{}. 
%This means that $\delta_{\rm s}$, the suppression factor of the generated lepton number asymmetry relative to the initial number density of AD field, is expected to be $\delta_{\rm s} \sim \mathcal{O}(10^{-3}-10^{-2})$.
%%the parametric dependence of the final baryon number asymmetry would be more-or-less same as the one in Ref.~\cite{}.

Since the  AD field is a flat direction during inflation, there is a potential danger of generating too large baryon isocurvature perturbations ($\mathcal{S}_{b \gamma} \equiv \delta \ln \l( n_B/s \r)$), but they are naturally suppressed in our scenario for reasons which will be explained below.

Setting $\langle \theta_\ell \rangle =0$ without loss of generality, from \eq{nL} one finds
\beq \label{d-nBs}
\delta \ln \l( \frac{n_B}{s} \r)
= -2 \frac{\delta X_{\rm osc}}{X_{\rm osc}} + 4 \l( \frac{\delta \ell_r}{\ell_0} + \cot(\Delta \theta) \frac{\delta \ell_a}{\ell_0} \r)
%\nonumber \\
%&\simeq& 8.9 \times 10^{-5} \l( \frac{m/\ell_0}{10^{-5}} \r) 
%\nonumber \\
%&& \times \l( e^{-\bar{\eta}_{\ell,0}^r N_{e,*}} + \cot(\Delta \theta) e^{-\bar{\eta}_{\ell,0}^a N_{e,*}} \r)
\eeq
where $\delta \ell_{r,a}$ is the radial/angular perturbations of $LH_u$ and we used $\delta \theta_\ell = \delta \ell_a / \ell_0$.
%, $\delta X_{\rm osc} = \delta \ell_{r,a}^0 = \bar{H}/(2 \pi)$, and $\bar{H} = m/\sqrt{3 \bar{\eta}}$ with $\bar{\eta} = 1.67 \times 10^{-2}$, and ignored the contribution from $\delta X_{\rm osc}$ since it is smaller than the other contributions by about two orders of magnitude.
%When an unstable flat direction is stabilized by a higher order operator, the physical mass squares of the radial and angular modes around the true vacuum are larger than the magnitude of the mass square parameter by a factor a few. 
%Hence, 
Denoting the $\eta$ of $LH_u$ around the true vacuum as $\eta_{\ell,0}^{r,a}$ for the radial and angular modes of $LH_u$, the perturbations of each mode at the end of inflation can be written as
\beq
\delta \ell_{r,a} \sim \delta \ell_{r,a}^0 e^{-\eta_{\ell,0}^{r,a} N_{e,*}}
\eeq
where $\delta \ell_{r,a}^0 = H/(2 \pi)$.
Note that $\eta_{\ell,0}^{r,a}$ can be larger than $|\eta_X|$ during inflation by a factor of a few, providing a suppression of $\delta \ell_{r,a}$.
Observations constrain the baryon isocurvature perturbations to be $\beta_{\rm iso} \equiv P_\mathcal{S}/\l(P_\mathcal{R} + P_\mathcal{S}\r) \lesssim 0.038$ with $P_\mathcal{S}$ being the power spectrum of the isocurvature perturbations \cite{Ade:2015lrj}.
This can be interpreted as
\beq
|\mathcal{S}_{b \gamma}| \lesssim \frac{\Omega_{\rm DM}}{\Omega_B} \l( \frac{\beta_{\rm iso}}{1-\beta_{\rm iso}} P_\mathcal{R} \r)^{1/2} \lesssim 5.34 \times 10^{-5}
\eeq
where $\Omega_{\rm DM}$ and $\Omega_B$ are the densities of cold dark matter and baryons, respectively.
For $A_\nu=\sqrt{|m_{LH_u}^2|}=m$ as an example, $\mathcal{S}_{b \gamma} = 2.37 \times 10^{-5} (m/\ell_0/10^{-5})$.
Therefore, we are safe as regards of  baryon isocurvature perturbations.

\subsection{Dark matter}
%If the axino (or the flatino) is the lightest supersymmetric particle (LSP) and the inflaton can decay to the LSPs, clearly the axino dark matter would overclose the universe unless the associated branching fraction is extremely suppressed which is unlikely.
%Therefore, the inflaton decay into axino need to  be kinematically forbidden. The parameter space required for this to happen was discussed in Ref.~\cite{Chun:2000jx}.
%Similarly, flaton decay to neutralinos should also be forbidden if the decay temperature of the flaton is lower than the freeze-out temperature of the neutralinos.

If the inflaton decays well after the freeze-out of the lightest neutralino which is likely to be our case, in order to avoid dark matter over-production, the inflaton decays into flatinos and neutralinos should be kinematically forbidden (for example, see Ref.~\cite{Chun:2000jx} for the parameter space).
Hence, dark matter in our scenario is likely to be either neutralinos from thermal bath or axions, depending on the choice of parameters and the initial conditions of inflation.

The energy density stored in radiation after inflation until the energy density of PQ-field ($\rho_{\rm PQ}$) becomes comparable to the brane tension is expected to be
\bea
\rho_{\rm r} &\sim& \frac{\Gamma_{\rm d}}{H} \rho_{\rm PQ} %\sim \sqrt{6} \Gamma_{\rm d} M_{\rm P} \Lambda^{1/2} 
\sim \sqrt{2} \l( \frac{\pi^2}{30} g_*(T_{\rm d}) \r)^{1/2} T_{\rm d}^2 \Lambda^{1/2}
%\nonumber \\
%&\sim& 4.6 \times 10^{-4} \l( \frac{T_{\rm d}}{10^{-2} m} \r)^2 \l( \frac{\Lambda^{1/4}}{m} \r)^2 m^4
\eea
where $g_*(T_{\rm d})$ is the number of relativistic degrees of freedom at inflaton decay.
Hence, the background temperature during this epoch ($T_\times$) is nearly constant irrespective of $H$, and it may or may not be lower than the typical freeze-out temperature of neutralinos in Einstein gravity, $T_{\rm fz} \sim m_\chi/20$ with $m_\chi$ being the mass of neutralino.

%It will be lower than the typical freeze-out temperature of neutralino ($T_{\rm fz}$), if 
%\bea
%\frac{T_{\rm d}}{T_{\rm fz}} 
%&\lesssim& \frac{1}{2^{1/4}} \l( \frac{\pi^2}{30} g_*(T_{\rm d}) \r)^{1/4} \l( \frac{g_*(T_{\rm fz})}{g_*(T_{\rm d})} \r)^{1/2} \frac{T_{\rm fz}}{\Lambda^{1/4}}
%\nonumber \\
%&\sim& \mathcal{O}(0.1)
%\eea
%Typically, one finds $T_{\rm fz} \sim m_\chi/20$.
%In addition, in the discussion of baryogenesis we found that $T_{\rm d}/m \lesssim \mathcal{O}(10^{-3}-10^{-2})$ may be necessary for a successful baryogenesis.
%This means $T_{\rm d}/T_{\rm fz} \lesssim \mathcal{O}(10^{-2} - 0.1)$ and the abundance of neutralinos is suppressed by a large exponential factor relative to the usual freeze-out abundance.

If $T_\times > T_{\rm fz}$, for $T_{\rm d} < T_{\rm fz}$ the abundance of neutralinos would be diluted by the entropy production generated in the decay of the inflaton.
%The would-be background temperature when $H \sim \Gamma_{\rm d}$ in case of no entropy production, denoted as $\tilde{T}_{\rm d}$, is
%\beq
%\tilde{T}_{\rm d} = T_{\rm fz} \l( \frac{H_{\rm d}}{H_{\rm fz}} \r)^{2/3}
%\eeq
%When $H \gg \Gamma_{\rm d}$, $\rho_{\rm r}(T_{\rm fz}) \approx \Gamma_{\rm d} H_{\rm fz} M_{\rm P}^2$ yielding
%\beq
%H_{\rm fz} \approx \l( \frac{\pi^2}{30} g_*(T_{\rm fz}) \r) \frac{T_{\rm fz}^4}{\Gamma_{\rm d} M_{\rm P}^2}
%\eeq
The dilution factor is given by
\beq
\Delta 
= \l( \frac{T_{\rm d}}{\tilde{T}_{\rm d}} \r)^3
= 3^{2/3} \l( \frac{g_*(T_{\rm fz})}{g_*(T_{\rm d})} \r)^{2/3} \l( \frac{T_{\rm fz}}{T_{\rm d}} \r)^5
%\nonumber \\
%&\simeq& 6.5 \times 10^{-7} \l( \frac{m_\chi}{T_{\rm d}} \r)^5
\eeq
where $\tilde{T}_{\rm d}$ is the would-be background temperature when $H \sim \Gamma_{\rm d}$ in case of no entropy production.
The relic density of neutralino can then be expressed as
%\bea
%\Omega_{\rm CDM} 
%&\sim& \Omega_{\rm CDM}^{\rm obs} \l( \frac{m_\chi}{100 \GeV} \r)^2 \frac{1}{\Delta}
%\nonumber \\
%&\simeq& 1.5 \times 10^{10} \Omega_{\rm CDM}^{\rm obs} \l( \frac{m_\chi}{10^4 \GeV} \r)^2 \l( \frac{T_{\rm d}}{m_\chi} \r)^5
%\eea
\beq
\frac{\Omega_{\rm DM}}{\Omega_{\rm DM}^{\rm obs}} 
\sim \l( \frac{m_\chi}{10^2 \GeV} \r)^2 \frac{1}{\Delta}
\sim 10^{10} \l( \frac{m_\chi}{10^4 \GeV} \r)^2 \l( \frac{T_{\rm d}}{m_\chi} \r)^5
\eeq
where we used $g_*(T_{\rm fz})=g_*(T_{\rm d})$ and $T_{\rm fz} = m_\chi/20$ for $\Delta$.
Hence, for $m_\chi \sim m_s \sim \mathcal{O}(10^{4-5}) \GeV$ the abundance of neutralinos can match the observed relic density of dark matter if $T_{\rm d}/m_\chi \sim \mathcal{O}(10^{-3}-10^{-2})$ which may also be compatible with that required for baryogenesis.

If $T_\times < T_{\rm fz}$, neutralinos would never be in thermal equilibrium and their abundance will be negligible, its exact size depending on $T_\times / T_{\rm fz}$. 
In this case, dark matter would have to be mainly axions.

As PQ-symmetry is broken during inflation, the axion domain wall problem is absent, and the contribution to dark matter comes  only from misalignment.
In this case, the axion coupling constant can be much larger than the typically allowed upper-bound, since the misalignment angle is given as the initial condition of inflation. 
%In this case \cite{}, 
%\beq
%\Omega_a h^2 = 0.11 \l( \frac{f_a}{5 \times 10^{11} \GeV} \r)^{1.184} \mathcal{F} \theta_a^2
%\eeq
%where $f_a = \phi_0/N_a$ is the axion coupling constant with $N_a$ being the anomaly coefficient of $U(1)_{\rm PQ}$ symmetry, $\mathcal{F}=\mathcal{F}(\theta_a) = \mathcal{O}(1)$ is the effect of anharmonicity, and $\theta_a$ is the misalignment angle at the onset of axion oscillation.

\section{Conclusions}

In this letter, we proposed an inflation scenario along a supersymmetric flat direction in the context of Randall-Sundrum type-II brane world.
Showing that, the mass scale ($m$) of the symmetry-breaking supersymmetric flat-direction is directly connected to the symmetry-breaking scale ($\phi_0$) as $m/\phi_0 \sim 10^{-7}$, 
we identified the Peccei-Quinn field associated with the axion solution to the strong CP-prblem as the natural candidate to play the role of the inflaton.
We also showed that a successful post-inflation cosmology including small active neutrino masses, baryogenesis and dark matter can be triggered by the non-zero vacuum expectation value of the PQ-inflaton via the generation of the heavy Majorna mass of the seesaw mechanism and the dynamical generation of the MSSM $\mu$-parameter at the end of inflation.
Either neutralinos or axions can be the main component of dark matter without an axion domain wall problem.
 
Our scenario does not have ad-hoc parameters except the brane tension ($\Lambda$).
The fundamental Planck scale in five-dimensional spacetime is expected to be of $\mathcal{O}(10^9) \GeV$ with a brane tension of $\mathcal{O}(10^{4-5}) \GeV$.
The brane tension imposes lower bounds on the scales of soft SUSY-breaking mass and breaking of $U(1)_{\rm PQ}$ symmetry at $\mathcal{O}(10) \TeV$ and $\mathcal{O}(10^{11}) \GeV$, respectively. 
This scenario can be excluded if primordial tensor perturbations are observed, a $\TeV$ scale SUSY (meaning $\Lambda^{1/4} < 1 \TeV$) is found, or the axion coupling constant turns out to be lower than $\mathcal{O}(10^{10}) \GeV$.  

In summary our scenario is a full description of the early history of the Universe from inflation to Big-Bang nucleosynthesis that may be easily falsifiable at future collider experiments.

\section{Acknowledgements}
The authors are indebted to Will Kinney for his helpful comments.
They also acknowledge support from the MEC and FEDER (EC) Grants SEV-2014-0398 and FPA2014-54459 and the Generalitat Valenciana under grant PROME- TEOII/2013/017.
G.B. acknowledges partial support from the European Union FP7 ITN INVISIBLES (Marie Curie Actions, PITN-GA-2011-289442).

\end{document}